\def\k{{\vec k}}
\def\q{{\vec q}} 

\def\a{\alpha}
\def\b{\beta}
\def\d{\delta}
\def\g{\gamma}
\def\e{\varepsilon}
\def\p{\pi}
\def\ph{\phi}
\def\s{\sigma}
\def\t{\tau}
\def\w{\omega}

\def\sumnext{\sum_{<i,j>}} 

\def\PR{{\rm Phys.\ Rev.}}
\def\PRL{{\rm Phys.\ Rev.\ Lett.}}
\def\JPSJ{{\rm J.\ Phys.\ Soc.\ Jpn.}}
\
\def\ni{\noindent}

\magnification=\magstep2
\def\la{ < \kern -12pt \lower 5pt \hbox{$\displaystyle \approx$}}
\def\ga{ > \kern -12pt \lower 5pt \hbox{$\displaystyle \approx$}}
\def\lsim{{ < \kern -11.2pt \lower 4.3pt \hbox{$\displaystyle \sim$}}}
\def\gsim{{ > \kern -11.2pt \lower 4.3pt \hbox{$\displaystyle \sim$}}}

\centerline{{\bf $g$-on Mean Field Theory of the $t$-$J$ Model}}
\vskip 5mm
\centerline{Flordivino Basco\footnote*{e-mail: basco@monet.phys.s.u-tokyo.ac.jp},  Hiroshi  Kohno,  Hidetoshi Fukuyama  
and  Ganapathy Baskaran$^1$}
\vskip 4mm
\centerline{Department of  Physics, University of Tokyo, 
7-3-1 Hongo, Bunkyo-ku, Tokyo 113} 
\centerline{$^1$ Institute of Advanced Studies, Princeton, NJ 08540,
USA\footnote{**}{Permanent Address: Institute of Mathematical Science, Madras 600 113, India}} \vskip 2mm
\centerline{(Received September 14, 1996)}
\vskip 7mm

\ni
{\bf Abstract }

	Implication of our recent proposal [$\JPSJ$ {\bf 65} (1996) 687] to
treat large-amplitude gauge-field fluctuations around the slave-boson
mean-field theory for the $t$-$J$ model has been explored in detail.  By
attaching gauge flux to spinons and holons and then treating them as free
$g$-on's which respect the time-reversal symmetry, the optimum
exclusion ($g$) and exchange ($\a$) statistics have been determined in the
plane of doping rate and temperature.  Two different relations between $\a$ 
and $g$ have been investigated, namely $g=|\a|$ (Case1) and
$g=|\a|(2-|\a|)$ (Case2).  The results indicate that 
slave fermion is favored at low doping while slave boson at high doping.
For two dimension, in Case1 intermediate statistics are found in between,
while in Case2 no intermediate statistics are found.  The consequences of
varying the dimensionality and strength of $J$ have been studied also.  The 
latter has no qualitative effect for both cases, while the former has a
profound effect in Case1.  

\vskip 4mm
\vskip 7mm
{\bf 1.  Introduction }

	Since the discovery of high-$T_{\rm c}$ cuprates$^{1)}$, there are
many attempts to understand this interesting class of materials.  It has
generally been accepted that the essence of electronic properties is in the
CuO$_2$ layers, and that the low-lying excitations of the layers are
described by a single-band $t$-$J$ model$^{2,3)}$.  Though the exact
results are not known, the slave-boson mean field theory$^{4-7)}$ has
played important roles in offering explicit predictions to be compared with
experiment as regards the doping dependence and effect of fermi surfaces,
and the agreement between the theoretical results $^{8-9)}$ and what have
been seen in the experiments on the spin excitations probed by neutron
scattering$^{10)}$ and nucler magnetic resonance (NMR)$^{11)}$,
phonons$^{12)}$ and transport properties$^{13)}$ are noteworthy.  The
scheme is based on spin-charge separation and the electron operator is
decoupled into spinon and holon, which is assumed to be fermion and boson,
respectively.  The phase diagram derived by the mean field approximation in
this scheme is given by Fig.1.  The states realized include
antiferromagnetic (AF) state, superconducting (SC) state, ``spin gap'' (SG)
phase, anomolous metallic (AM) state, and electron liquid (EL) state.  The
AM state, where both spinons and holons move coherently and essentially
independent of each other, i.e. spin and charge are separated, is described
as the uniform RVB ($u$-RVB) state.  In the SG state, where the spin and
charge are also separated, the spinons make short range singlet pairs,
which is described as the singlet RVB ($s$-RVB) state.  On the other hand
in the EL state spinons and holons are no longer independent, because of
the bose condensation of holons, and coupled together to form electrons.
Based on this phase diagram various physical quantities especially, the
spin gap phenomenon typically seen in the rate of NMR$^{11)}$ have been
discussed$^{8-9)}$ in connection with the SG state in the phase diagram.  

	It is generally the case, however, there exist fluctuations around
the mean field solution.  The most important fluctuations, in the present
case, are expressed by the phases of the order parameters in the $u$-RVB
state, which are viewed as U(1) gauge fields$^{14)}$.  The charge transport
properties governed by holons are essentially influenced by these gauge
fields because of the small characteristic energy of holons.  The linear
$T$-dependence of the resistivity has been discussed by treating the gauge
field in the Gaussian approximation$^{15)}$.  However there are indications
that the gauge-field fluctuations  can be large.  For example, high
temperature expansion on the lattice$^{16)}$ indicates that for a
reasonable value of $t/J$ the root-mean-square of fluctuations of gauge
flux is large ($\sim \pi$ per plaquette) .  This might lead to
quantitative$^{17)}$ and even qualitative$^{18)}$ changes in the mean field
phase diagram. The existence of such large fluctuations may indicate the
necessity to elaborate the description of the ground state, and there are
some efforts towards this target.  On this line, there are two approaches
both of which try to incorporate the fact that in the limit of vanishing
doping rate, i.e. half-filling, the $\p$-flux state is the most stable
state.  One idea exploits the equivalence of the $\p$-flux state to
$d$-wave due to SU(2) symmetry$^{19,20)}$, while another maps the $\p$-flux
state into slave semion scheme$^{21)}$ by binding fluxes of half-quanta to
both spinons and holons.  The latter type of approach to treat strongly
correlated electron systems may date back to the work of Kalmeyer and
Laughlin $^{22)}$ who mapped the two-dimensional spin system to the
fractional quantum Hall system. However, since flux state is stable only at
half-filling, it is an open question as regards what will happen in finite
doping rate.

	 Recently, we have proposed a different approach$^{23)}$ to treat
these large gauge-field fluctuations.  The spinons and holons may avoid
these large gauge-field fluctuations by partly absorbing them, which
results in statistical transmutation of these particles.  The statistics of
holons and spinons will be chosen such that the free energy is minimum,
i.e. the statistics are to be optimized.  Here the time-reversal symmetry
is properly taken into account  by considering the fluctuating nature of
the gauge field.  A similar problem of determining variationally the
optimum statistics of spinons and holons has been studied also by
Mori$^{24)}$ for the case of the Hubbard model with infinite Coulomb
interaction.  His formalism is, however, limited to continuum space, breaks
time reversal symmetry and it turns out that only slave boson is realized
at all doping rates.    

	In our explicit calculations of the thermodynamic properties the
exclusion properties of anyons (particles with flux) has been noted.  This
exclusion property is based on the idea of Haldane$^{25)}$ which
generalizes the concept of Pauli exclusion principle.  A particle obeying
this exclusion principle is called ``$g$-on''$^{26)}$ or
``excluson''$^{27)}$.  Its distribution function was determined by
Wu.$^{28)}$  Recently, some models are known to realize
$g$-on's$^{29,30)}$.    

	This paper has the following outline.  Section 2 describes our
model and formulation.  The results of our numerical calculation are given
in Sec.3.   In Sec.4 conclusions are given together with discussions on the 
possible relevance in describing the anomalous metallic states of
high-T$_c$ cuprates.  A part of the results for two dimensions in this study
have been published elsewhere$^{23)}$.

\vskip 4mm
{\bf 2.  Model and Formulation }
\vskip 1mm
\noindent
{\it 2.1)  Slave Boson Mean-Field Theory}
\vskip 4mm

	The $t$-$J$ model is defined by the Hamiltonian,
$$\eqalignno{   
	H = - & \sum_{i,j,\s} \bigl[t_{ij} (1-n_{i-\s}) c^\dagger_{i\s} c_{j\s}(1-n_{j-\s})  + {\rm h.c.} \bigr] \cr
	+  & J\sumnext {\bf S}_i  {\bf S}_j - \mu_e \sum_{i,\s}n_{i \s}  .
     &(2.1)} 
$$
where the parameters and operators, especially, ${\bf S}_i = {1\over 2}
c^\dagger_{i\a} {\bar {\t _{\a \b}}} c_{i\b}$ with ${\bar \t}$ being Pauli's spin
matrices,  have the usual definitions, and $<,>$ means the summation over
pairs of nearest neighbor sites.   

	In slave boson scheme the electron operator is decomposed  in terms
of holon 
($h_i$) and spinon ($s _{i\s}$), $c_{i\s } = h^\dagger_i s_{i\s }$.  The
holons and spinons are assumed as bosons and fermions, respectively.  The
Hamiltonian is rewritten as 
$$\eqalignno{   
H = - & \sum_{i,j,\s} \bigl[ t_{ij} h_i s^\dagger_{i\s} s_{j\s} h^\dagger_j
+ {\rm h.c.} \bigr]  + J \sumnext {\bf S}_i  {\bf S}_j  \cr
 - & \mu_e \sum_{i,\s}s^\dagger_{i\s} s_{i\s}  - \sum_i \lambda_i \bigl(\sum_\s
s^\dagger_{i\s} s_{i\s} + h^\dagger_i - 1 \bigr)   
&(2.2)} 
$$
where the last term with Lagrange multiplier, $\lambda_i$, excludes 
double occupancy at any sites.   In the mean field approximation the local
constraint is being relaxed to global one, $\lambda_i \longrightarrow
\lambda$.  It has been shown$^{8)}$ that the choice of values of the
transfer integrals, ($t/J=4$) $t'/t=-{1\over 6}$, $t''/t=0$ and 
$t'/t=-{1\over 6}$, $t''/t={1\over 5}$  simulate the fermi surfaces of
LSCO and YBCO, respectively.  Here $t$, $t'$, and $t''$ are transfer
integrals between the  first
nearest neighbors, the next nearest neighbors, and the 3rd nearest neighbors,
respectively. For simplicity our work will consider only the standard
$t$-$J$ ($t/J=4$ and $t'/t=t''/t=0$) since the present 
conclusions are expected to be insensitive to these choices.  Moreover,
although the mean field theory predicts  various states as shown in Fig.1
we will limit ourselves to the $u$-RVB state, where $\chi_s=<\sum_\s
s^\dagger_{j\s } s_{i\s }>$ and $ \chi_h=<h_i h^\dagger_j>$ are the order
parameters.  The effect of singlet spinon pairing will be studied in a
separate publication. Then, the mean field hamiltonian, $H_{MF}$,  is    
$$\eqalignno{   
H_{MF} = -  & \sum_{<i,j>,\s} \bigl[ t_{ij}({\rm s}) s^\dagger_{i\s} s_{j\s}
+ {\rm h.c.} \bigr] - \mu_s \sum_{i,\s} s^\dagger_{i\s}   s_{i\s} \cr  
         - & \sumnext \bigl[t_{ij}({\rm h})h^\dagger_i  h_j + {\rm h.c.}
\bigr]   - \mu_h \sum_i h^\dagger_i  h_i 
	&(2.3)} 
$$
where $\mu_h \equiv \lambda$, $\mu_s = \mu_e + \mu_h$ are the
chemical potentials of holons and spinons, respectively, and $t_{ij}({\rm h})
= t  \chi_s$, $t_{ij}({\rm s})= t  \chi_h + 3J \chi_s / 8$ are the
effective hopping integrals of holons and spinons.

\vskip 4mm
\noindent
{\it 2.2)  Fluctuations around the Mean-Field Solution}
\vskip 4mm

	The original slave boson scheme without mean field approximation
the electron operator, $c_{i\s } = h^\dagger_i s_{i\s }$, is invariant
under local phase transformation which is lost totally in the mean field
approximation.  As Baskaran and Anderson$^{14)}$ noted, to recover this
local symmetry, the fluctuations around the mean field solutions should be
treated as U(1) gauge field and hence the order parameters are not
constants but fluctuating in space and time; $t_{ij}(q) = t_q {e^{i
a_{ij}}}$ ($q=h,s$) where $t_q$ are the mean field values. The
fluctuation ${a_{ij}}$ is, indeed, a gauge field that affects the spinons
and holons together.  This cures the complete decoupling of the electron
operator at the mean field level.  The Hamiltonian will then be written as 
$$\eqalignno{   
H = -  & t_{\rm s} \sum_{<i,j>,\s} \bigl [{e^{i {a}_{ij}}}
s^\dagger_{i\s} s_{j\s}  + {\rm h.c.} \bigr] - \mu_s \sum_{i,\s}
s^\dagger_{i\s}   s_{i\s}  \cr 
         - & t_{\rm h} \sumnext \bigl[ {e^{i {a}_{ij}}} h^\dagger_i  h_j +
{\rm h.c.} \bigr]  - \mu_h \sum_i h^\dagger_i  h_i . 
     &(2.4)} 
$$
This is invariant under the following gauge transformation, 
$$\eqalignno{   
	s^\dagger_{i\s}& \longrightarrow s^\dagger_{i\s}  {e^{-i \ph _i}} , \cr
	h_i &\longrightarrow h_i  {e^{i \ph _i}} ,  \cr
	{a}_{ij}  &\longrightarrow {a}_{ij} + (\ph _i - \ph _j) .  
&(2.5)} 
$$

	In the calculation of transport coefficients$^{15)}$ the
fluctuations have been treated in the Gaussian approximation and it has
been indicated that gauge fluctuations are not small.
 
	Here we propose a way to take account of these large gauge-field
fluctuations by assuming that the fluctuating gauge fluxes are
approximately of Ising type, i.e. the magnitude is nearly constant but its
sign is changing spatially and temporally [Fig.2(a)].  In this paper we
will determine the magnitude of these Ising-type fluctuations that optimize
the free energy by exploiting the ideas of flux binding and exclusion
statistics.  
 
\vskip 4mm
\noindent
{\it 2.3)  Flux Binding and Exclusion Statistics}
\vskip 4mm

	In a region where gauge-field fluctuations have some fixed value and
then the uniform (internal) magnetic field is present, the fluxes due to
gauge-field fluctuations will be attached to  spinons and holons
[Fig.2(b)].  Now we have a system of anyons whose exchange statistics $\a$
(defined as the phase factor ${{\rm e}^{i\a \p}}$ accumulated by the total
wave function when two particles are interchanged) has approximately a
constant absolute value.  The new particles are then
defined as $\tilde s_{i\s} = s_{i\s} {\rm e}^{i\alpha Q_i} $ and $
\tilde h_i = h_i {\rm e}^{i\alpha Q_i}$.  Here $ Q_i = \sum_{j\ne i}
\theta_i(j) (h^\dagger_j h_j + \sum_\s s^\dagger_{j\s} s_{j\s}$) and
$\theta_i(j)$ represents the angle of a vector from site $j$ to $i$
with respect to some fixed reference axis.  It is straightforward to see  
$$
\tilde h_i \tilde h^{\dagger}_j = e^{i\a \p} \tilde h^{\dagger}_j \tilde h_i, 
 \hskip 1cm
\tilde h_i \tilde h_j = e^{i\a \p} \tilde h_j \tilde h_i, \hskip 1cm i\neq
j
$$
and
$$
\tilde s_{i\s} \tilde s^{\dagger}_{j\s} = - e^{i\a \p} \tilde
s^{\dagger}_{j\s} \tilde s_{i\s} , 
\hskip 1cm
\tilde {s_{i\s}} \tilde {s_{j\s}} = - e^{i\a \p} \tilde {s_{j\s}}  
\tilde {s_{i\s}}, 
\hskip 1cm i\neq j 
\eqno (2.6)
$$
and then the exchange statistics of holons and spinons are $\a$ and $\pm 1 +
\a$, respectively. The onsite commutation relations are simply given by the 
commutation relations of the undressed particles.  The Hamiltonian for
$\tilde h_i$ and  $\tilde s_{i \s}$ is the same as Eq.(2.4) but we expect
that the magnitude of fluctuations $\tilde a_{ij} = a_{ij} - \a (Q_i -
Q_j)$ are much reduced.  Then we assume that the fluctuations $\tilde
a_{ij}$ felt by the new particles  $\tilde h_i$ and  $\tilde s_{i \s}$ are
negligibly small and the mean field approximation is meaningful.  If we fix
$\alpha$, it will break the time-reversal symmetry.
However, in the present case the time-reversal symmetry is maintained due
to the  fluctuating nature of $\a$; the magnitude is fixed but the sign is
fluctuating as shown in Fig.2(c), so that the (net) flux will vanish
as an average and the thermodynamic properties of the system will be
purely described by the exclusion properties by the dressed holons and
spinons.  In other words, although the system is composed of domains with
fluxes $\a$ and $-\a$, it will be regarded as homogeneous in its exclusion
properties if we neglect domain boundaries [Fig.2(d)], and its
thermodynamics may be approximately treated by use of the $g$-on
distribution function$^{28)}$ 
$$
n_g(\e) = (w_g(\e) + g)^{-1} .
$$
Here $w_g(\varepsilon )\equiv w$ is determined by   
$$ 
 w^g (1+w)^{1-g} = {\rm e}^{\beta (\e - \mu) }, \eqno(2.7) 
$$  
 with one-particle energy $\varepsilon$, chemical potential $\mu$ and
inverse temperature $\beta$. For a special value of $g=0$ ($g=1$) we have
the familiar bose (fermi) distribution function.  The distribution
functions for several choices of $g$ are shown in Fig.3.  Here the
exclusion statistics is defined as the decrease in the size of one-particle
Hilbert space as the number of particles increases$^{25)}$.  For boson this
size does not change, $g=0$, and for fermion the decrease is one whenever
the number of particles increases by one, $g=1$.  In between boson and
fermion we have $g$-on, $0<g<1$.  

	There are so far two proposed relations between the exclusion
statistics $g(\geq 0)$ and exchange statistics $\alpha$ ($-1\leq \alpha
\leq 1$); namely, 
$$
Case1: \hskip 2mm g=|\alpha|  \eqno(2.8a)
$$
and
$$
\hskip 15mm Case2: \hskip 2mm g=|\alpha| ( 2- |\a|) . \eqno(2.8b) 
$$
These relations, Eqs.(2.8a) and (2.8b), are the exclusion
statistics of system of free-anyon with Dirac-type dispersion$^{31)}$ and
with quadratic dispersion$^{32)}$, respectively, which are classified as
Case1 and Case2 in the following.  In the present study to see overall
tendency as the doping rate is varied, the energy spectra of spinons and
holons vary in a wide range and the correct assignment is not known and
then we simply assume either (2.8a) or (2.8b) and investigate the
consequences for both cases.  If we assign the exclusion statistics
$g_h=g(\a)$ for holon, then $g_s=g(1-\a)$ for spinon since they are
composite particles of electron. Hence the free energy becomes a  function
of $\d$ (doping rate), $T$ (temperature) and $\a$ (statistics).  By
treating $\a (0\leq \a \leq 1)$ as a variational parameter we can minimize
the free energy and determine the phase diagram in the plane of $\d$ and
$T$.  In this study the dimensionality, $d$, of the system can be arbitrary
and we will examine the cases of $d = 1, 2, 3$.

	The free energy is given by
$$\eqalignno{   
F = -2T & \sum_\k {\rm ln} (1+w^{-1}_{g_s}(\e_\k))
         - T\sum_\k {\rm ln} (1+w^{-1}_{g_h}(\w_\k)) \cr
   + N & \bigl[ d \big(2 t\chi_s\chi_h + {3J\over 8}\chi^2_s \big) + (1-\d
)\mu_s + \d \mu_h \bigr] 
	&(2.9)} 
$$
and the self-consistent equations are 
$$\eqalignno{   
\d = & {1\over N} {\sum}_k n_{g_h} (\w_\k) , \hskip 2cm
{\d}_s \equiv {1\over 2} (1 - \d) = {1\over N} {\sum}_k n_{g_s} (\e_\k ) , \cr
\chi_h = & {1\over {dN}} {\sum}_k \g_\k  n_{g_h} (\w_\k)  , 
\hskip 4mm {\rm and} \hskip 4mm
{\chi}_s = {2\over {dN}} {\sum}_k \g_\k  n_{g_s} (\e_\k ) ,
	&(2.10)} 
$$
where  $ \e_\k = -2t_s \g_\k - \mu_s$,  $ \w_\k = -2t_h \g_\k - \mu_h $  
with $ \g_\k = {\sum^d  _{x=1}} \cos k_x$, $ n_{g_q}$ ($q=h,s$) the
``$g$-on'' distribution function and $N$ is the total number of lattice
sites.    

\vskip 4mm
{\bf 3.  Results}
\vskip 4mm
	In the following two subsections the dimensionality of the system
is fixed to two, $d=2$, whereas effects of the dimesionality will be
discussed afterwards.
\vskip 4mm

\noindent
{\it 3.1) $T=0$}
\vskip 4mm

	First we will consider the $T=0$ where the physics and the 
overall features of the results are transparent and easy to comprehend.
In this case the distribution function is a step function with the
``$g$-$on$ic fermi energy'' $\mu_q$, $n_{g_q} ={g_q}^{-1} \theta(\mu_q -
\e^q)$ ($q=h,s$) where  $\e^h= \w_\k$ and $\e^s=\e_\k$.  The energy is 
$$
	E = - N d \bigl( 2 t\chi_s\chi_h + {3J\over 8}\chi^2_s \bigr)
  \eqno (3.1) 
$$
where the order parameters,  $\chi_q$ ($\q=h,s$), are implicit functions of
the statistics, $\a$.  In Figs.4 $(a)$ and $(b)$, the energy per lattice
site as a function of $\a$ for several values of doping rate are shown for
Case1 and Case2.  One can easily see that for both cases slave fermion 
($\a=1$) is favored for low doping region and slave boson ($\a=0$) for high
doping region.  However there exists a difference.  In Case1 the optimum
statistics changes continuously from slave fermion to slave boson through
the intermediate statistics ($0<\a<1$, to be referred to as slave $g$-on in
the following) for $0.351 < \d < 0.392$, as seen in Fig.5(a).  On each
boundary of two different statistics the transition is of second order.  On
the other hand, in Case2, the optimum statistics changes from $\a=1$ to $\a=0$
discontinuously at $\d=0.371$, a first order transition as seen in
Fig.5(b).  This first order transition is accompanied by a discontinuous
change of the fermi surfaces of holons and spinons.  
	
	The result can be understood as due to bose-condensation of
bosons.  In the low doping region, spinons are abundant and govern the
energetics; the energy gain is maximum if all spinons are bosons and
condense into the bottom of the band. Hence, slave fermion ($\a=1$) is
favored at low doping as also has been shown in Ref. 33.  On the other
hand, in high doping region holons are abundant and the maximum energy gain
occurs if they become bosons and condense and slave boson ($\a=0$) is favored.     

\vskip 4mm
\noindent
{\it 3.2) Finite Temperature}
\vskip 4mm

	Let us first study high temperature behaviours.  The onset
temperature of the $u$-RVB state is 
$$
T_D = {3\over 8} J a_s \bigl[1 + \sqrt {1 + 2 {a_h\over a_s} ({8t\over
{3J}})^2} \bigr] ,   \eqno (3.2) 
$$
where $a_q = \d _q (1-\d _q g_q) (1+\d _q {\bar g_q})$, ${\bar
g_q} = 1 - g_q$ , $(q=h,s)$, $\d_s={1\over 2}(1-\d)$ and $\d_h \equiv \d$
.  This expression is derived by assuming that the transition is of second 
order, and does not hold when the transition is of first order (see \S
3.3, and \S 3.4).  
 
	Above this temperature the effective hopping integral or bandwidth is
zero and there is no energy gain due to itineracy.  Then, the free energy
for $T>T_D$
is determined purely by the entropy,
$$
S = N \big\{ \d {\rm ln} \d + (1 - \d g_h) {\rm ln}  (1 - \d g_h) -  (1 + \d
	\bar{g_h}) {\rm ln} (1 + \d  \bar{g_h})
$$
$$
  + 2 \bigl[ \d_s {\rm ln} \d_s + (1 - \d_s g_s) {\rm ln}  (1 - \d_s g_s) -
   (1 + \d_s \bar{g_s}) {\rm ln} (1 + \d_s \bar{g_s}) \bigr] \big\} .
  \eqno (3.3) 
$$
Fig.6(a) and (b) show $-S/N=F/NT$ (free energy per lattice site divided by
temperature) for different values of doping rate for Case1 and Case2,
respectively.  These fix the feature of the phase diagram at $T>T_D$;
the intermediate statistics region is possible for $0.371<\d<0.457$ in
Case1 but not possible in Case2.  We remark that this is independent of 
spatial dimensionality, since Eq.(3.3) is.  As the temperature is
lowered through $T_D$, the transition to $u$-RVB state is of second order
for both cases, except for a small region of doping $0.4043<\d<0.4124$
(from A to B of Fig.7(b)) in Case2 where the transition is of first order.
This is because the $T_D$'s on both sides of slave fermion-slave boson
interface do not fall on the same $\d$, i.e. the $T_D$ from A to B
coincides with the interface of slave fermion and slave boson [see inset of
Fig.7(b)].  The overall phase diagram for Case1 and Case2 are shown in
Fig.7(a) and (b), respectively.   
  
	In Case1, the width and the location (in $\d$-space) of the region
where the slave $g$-on is stabilized, is almost independent of $T$ because
$-S/N$, as a function of $\a$, has basically the same features as $E/N$ for
the same range of doping $\d$.   

	In general, the absence or presence of intermediate statistics can
be understood in terms of the total derivative of the free energy with
respect to $\a$, 
$$ 
{dF(\a)\over {d\a}} = T \bigl[g'(\a) f_{g_h} - 2  g'(1-\a) f_{g_s}\bigr]
  \eqno (3.4) 
$$
where  $f_{g} = \sum_\k n_{g} {\rm ln}(1+w^{-1}_{g})$, ($g=g_h,g_s$) and
the prime denotes 
differentiation with respect to its argument.  For Case1, at low
doping 
$dF(\a)\over {d\a}$ is negative while at high doping it is positive
for any  
value of $\a$. This is due to fact  $f_{g_s}>f_{g_h}$ ($f_{g_h}>f_{g_s}$)
for low (high) doping.  At the same time it is seen that
${dF(0)\over{d\a}}<0$ and  ${dF(1)\over {d\a}}>0$ for some range of $\d$.
  With this we note that there exist a region of $\d$ where intermediate
statistics is possible.   
 
	For Case2, on the other hand, it is straightforward to see that
${dF(\a)\over {d\a}}$ has only one zero for $0\leq \a \leq 1$ and
${dF(0)\over {d\a}}> 0 $ and ${dF(1)\over {d\a}} < 0$, thus, intermediate
statistics is impossible to occur in this case.  

	We also considered another hypothetical assignment of statistics
$g=\a^2$ where at $\a=0$ and $\a=1$ reduce properly to boson and fermion,
respectively.  In this assignment, a wide region of intermediate statistics
was  found because ${dF(0)\over {d\a}}<0$ and ${dF(1)\over {d\a}}>0$ .  

	The possibility of intermediate statistics may generally be
understood by  an effective exclusion statistics $g_{eff}$ defined by
$g_{eff} = g_h + g_s$.  If $g_{eff} > 1$, it is impossible to
realize, whereas, if $g_{eff} < 1$, the possibility of its occurrence is very
high.  The case of $g_{eff}=1$, which corresponds to Case1, is marginal and
elaborate analysis is needed. 

\vskip 4mm
\noindent
{\it 3.3) Effect of Dimensionality }
\vskip 4mm	

	The $g$-on as a generalization of boson and fermion is not limited
to two dimensional in contrast to anyon whose realization is
restricted to $d=2$.  In this subsection we  extend our calculations to
$1d$ and $3d$ under the assumption that the free energy given by Eq.(2.9)
is still valid in these cases and explores its consequences  that may be
relevant in understanding more on the '$g$-on physics' for the $t$-$J$
model.   

	In this subsection only Case1 will be discussed  where the role of 
dimensionality is profound in contrast with Case2 which is insensitive to
the change of dimension.  The phase diagrams for $1d$ and $3d$ are shown 
in Fig.8(a) and (b), respectively, where the bose condensation temperature,
$T_B$, is also drawn for $3d$.  The main effect of increasing the
dimensionality is to reduce the slave $g$-on region as seen in Figs.8(a),
7(a), and 8(b) for $1d$, $2d$, and $3d$, respectively.  This is also seen
in Fig.8(d) where, at $T=0$, the optimum statistics as a function of $\d$
for each dimension is shown.  The energy as a function of $\a$ for
different values of $\d$ in each dimension further supports these
understanding as seen in Figs. 9(a)($1d$), 4(a)($2d$) and 9(b)($3d$]. 
Clearly, this suppresion of slave $g$-on-phase as the dimension increases
is due to increasing tendency of bose condensation.   

	In $1d$, at $T=0$, the entire region of $\d$ favors slave
$g$-on-phase except at the endpoints, $\d=0$ and $\d=1$ where slave fermion
and slave boson, respectively, are realized.  On both ends the transitions
are of second order.  However, the deviation of optimum statistics from
$\a=1$ near $\d=0$ and from $\a=0$ near $\d=1$ are very small as seen in
Fig.8(c).  At finite temperature, in both regions of doping rate the
change of statistics are continuous also.  However, even at very low
temperature the region of $g$-on is reduced appreciably compared to that at 
$T=0$ as seen in Fig.8(a).  This is due to singularity of the density of
state at the band bottom which make the effect of entropy important even
at very low temperature.  
 
	In $3d$, on the other hand slave boson and slave fermion phases are 
stabilized much compared to $2d$.  This is due to bose condensation.  The
slave $g$-on phase is only realized at temperature near and above $T_D$
where the effect of entropy is significant.  We note there exists a  triple
point ${\rm P_{tr}}$($T_{tr}=1.732J,\d_{tr}$=0.373) where the three phases
(slave fermion, slave $g$-on, and slave boson) coexist, at relatively high
temperature [Fig.8(b)].  This is because $T_B$ scales with the band
width and has the same order of magnitude as $T_D$.  Below ${\rm P_{tr}}$,
the transition at fixed temperature from slave fermion to slave boson is of
first order.  At ${\rm P_{tr}}$, the slave fermion-slave boson interface
branches into slave fermion-slave $g$-on and slave $g$-on-slave boson
interfaces.  It turns out that the transition from slave fermion to slave
$g$-on phases (line ${\rm P_{cr}Q}$) is of second order, while the
transition between the slave $g$-on and slave boson is of first order on
the line  ${\rm P_{tr}P_{cr}}$ but becomes of the second order on  ${\rm
P_{cr}R}$.  In Fig.8(c), the $\a$-dependece of the free energy are shown at 
X, Y, Z (inset) where X and Y are points in slave fermion-slave boson
boundary and Z is inside the slave $g$-on region which is chosen such that
the free energies of slave fermion and slave boson are the same. 

	Another consequence of $T_B\lsim T_D$ is that bose condensation of
spinons becomes of first order for $\d<0.0530$.  This region is indicated
by the dotted line in Fig.8(b).  Moreover, we found numerically that for
very small $\d$ ($\leq 0.0062$), the transition into the $u$-RVB state
occurs at higher temperature than that given by Eq.(3.2), which has been
derived by assuming second-order transition.  What actually happens is that 
$u$-RVB is driven by bose condensation of spinons, namely, $T_D$ coincides
with $T_B$, and becomes of first order.  This is seen from lines $1$ and
$2$ in Fig.10(a), where the Ginzburg-Landau free energy, $\d F \equiv
F$($\chi_s$)$-F$($\chi_s=0$), as a function of $\chi_s$ has two
minima.  Note that the bose condensation is stabilized for $\chi_s$ larger
than the value indicated by dot in each line.    At slightly larger
$\d$($\gsim 0.0062$), the $u$-RVB transition becomes of second order but
bose condensation remains a first order transition (lines 3-7 in
Fig.10(a)), until $\d= 0.0530$ beyond  which $T_B$ becomes of second order.
The phase diagram  very close to half filling is shown in Fig.10(b).   

	Hence, as an overall view the richness of our phase diagrams
discloses the intricate interplay between bose condensation, band edge
singularity of the density of states, and entropy at high
temperature. In $1d$, entropy is the key factor, except at $T=0$.
However, in $3d$ bose condensation is the
key factor.  In $2d$, the intermediate statistics are realized on the same
range of both high and low doping and then we may infer
that the tendency to bose condense and the effect of entropy have the
same weight. 

\vskip 4mm
\noindent
{\it 3.4) Effect of $J$ }
\vskip 4mm	
  
	We next study the effect of $J$.  The global features of our
results do not depend on the magnitude of $J$ .  For example, the phase
diagram for $J=0$ in $2d$ is shown by the dotted lines in Fig.7 and do
not deviate much from the results of finite values of $J$.  The same is
true for $1d$ and $3d$.  However in $3d$ for $J=0$, the transitions at
$T_D$ and $T_B$ are both of second order at low doping in contrast to the
case of finite $J$.  Moreover, in Case1 (Case2) the $T_B$ starting from the
interface 
of slave $g$-on(slave fermion)-slave boson  phases up to $\d=0.6439$ is of
first order transition.  Qualitatively, the main effects of $J>0$ are
pronounced at low doping region where the number of spinons are large.
For example at half-filling, it removes the degenaracy in the optimum
statistics at $T=0$ and supports finite $T_D$, as well as $T_B$ in $3d$.
Furthermore, it  stabilizes the slave fermion-phase and, in general, raises
$T_D$ especially at low doping region.  In Case1, the former results in the
narrowing of the region of the intermediate-statistics phase.  All these
can be understood as the result of increase in the coherency of spinons,
and consequently of holons, due to $J$.    

\vskip 2mm
\noindent
{\bf 4.  Summary and Discussions}
\vskip 4mm
	
	In the slave-boson theory of the $t$-$J$ model, the electron
operator is decomposed into spinons and holons, which are assumed to be
fermions and bosons, respectively.  Within the mean field approximation,
the spinons and holons are assumed to move independently and the
characteristic phase diagram is predicted as given by Fig.1.  The most
important fluctuations around this mean-field solution are the gauge
fields, which by themselves couple to spinons and holons leading to the
effective interaction between them.  Especially once the bose-condensation
of holons is realized, i.e. $T<T_B$, these effects of gauge fields are
essential resulting in the binding of spinon and holon which implies that
the ordinary view based on electrons is recovered.  In the region where the 
spin and charge are separated, i.e. in the region of anomalous metal (AM)
and spin gap (SG) in Fig.1, various predictions have been made for spin
excitations based on the mean-field approximation without the explicit
recourse to the fluctuation effects, which are expected to be relatively
minor because of large characteristic energy due to Fermi degeneracy of
spinons.  Above all the possibility has been pursued$^{8,9)}$ of the
identification 
of the spin gap phenomenon with the onset of the SG phase out of the
anomalous metallic (AM) state, the former and the latter are singlet and
uniform RVB states, respectively.  This identification has been first
suggested by Rice$^{34)}$ who payed attention to the quantitatively different
temperature dependences of the NMR shift and rate in the optimal and
underdoped YBCO, and this fact has been further explored in Ref. 35.  This
identification has led to the prediction of the 
anomalous frequency shift of the particular in-plane oxygen phonon
modes$^{9)}$, which has been confirmed experimentally$^{12)}$.  The results
of the recent angle-resolved phtoemission spectroscopy seem to be also in
accordance with the prediction$^{36)}$. 

	In contrast to these spin excitations, the charge transport
properties, i.e. the resistivity and the Hall effect coefficient, are very
susceptible to the fluctuations around the mean field solutions, since the
holons are assumed to be  pure boson.  Moreover there are theoretical
indications that these fluctuations can be very large.

	The present study is intended to overcome this apparent difficulty
by proposing a new theoretical framework to treat the large gauge
fluctuations.  By attaching the gauge fluxes to spinons and holons, which
are assumed to have a fixed value but with the fluctuating signs (like
Ising model), and then treating these particles as free $g$-on's, the
statistics has been optimized by minimizing the free energy on the plane of
doping rate and the temperature by use of the distribution functions of
$g$-on's.  (The brief report of the present study has been given in
Ref.23.)  To best of our knowledge, this is the first theory to search for 
statistics  variationally and moreover this theory repects the
time-reversal symmetry.  The result of the optimization is that the slave
fermion, where spinons are bosons, is favored at low doping (near
half-filling) while slave boson, where spinons are fermions, is favored at
high doping, with the intermediate statistics in between.  This general
feature of slave-fermions and slave-bosons are easily understood by the
existence of the bose condensation of spinons at low doping and holons at
high doping.  This understanding also clearly indicates that the regions of 
pure slave-fermions or slave-bosons are overemphasized in the present
approximation because the existence of hard cores of spinons and holons is
totally ignored.  Consequently we expect the region of the intermediate
statistics will be enlarged in a more elaborate treatment.  In such a
region of intermediate statistics the gauge fluctuations are greatly
reduced because of the existence of the finite mass in the gauge
propagators due to the Chern-Simons term$^{37)}$.  It is possible that such
a region of intermediate statistics (but close to the slave-boson) is the
one where many interesting and anomalous properties have been observed
experimentally.   

	At the same time we also note the following questions which still
remain to be addressed properly.  The assignment of statistics, $g=g(\a)$,
we adopted in the present study is derived for anyons in continuum space 
and does not reflect the actual dispersion of our spinons and holons.
Moreover, the assignment was been made by comparing the second virial
coefficients which are appropriate only at high temperature.  Hence,
further investigations are needed on the lattice effects as well as the
temperature dependences. 

\vskip 1cm
\noindent
{\bf Acknowledgement} 
\par 
 	G.B. is thankful to JSPS for enabling him to stay in the Department
of Physics, University of Tokyo.  F.B., on leave from the University of the
Philippines, is grateful to the Hitachi Scholarship Foundation for
financial support.  This work is financially 
supported by a Grant-in-Aid for Scientific Research on Priority Area \lq\lq
Anomalous Metallic State near the Mott Transition" (07237102) from the
Ministry of Education, Science, Sports and Culture.  

\vskip 1cm
\noindent
{\bf References}
\vskip 3mm
\item{1)} G. Bednorz and K.A. M\"{u}ller: {\rm Z. Phys.} {\bf B 64} (1986)
189. 

\item{2)} P.W. Anderson: Science {\bf 235} (1987) 1196.

\item{3)} F.C. Zhang and T.M. Rice: $\PR$ {\bf B37} (1988) 3759.

\item{4)} G. Baskaran, Z. Zou and P.W. Anderson: Solid State Commun. {\bf 63} 
	(1987) 973 .  

\item{5)} Y. Suzumura, Y. Hasegawa and H. Fukuyama: $\JPSJ$ {\bf 57} (1988)
	2768; Physica {\bf C 153-155} (1988) 1630.

\item{6)} I. Affleck and J.B. Marston: $\PR$ {\bf B37} (1988) 3774. 
 
\item{7)} G. Kotliar and J. Liu: $\PR$ {\bf B38} (1988) 5142. 

\item{8)} T. Tanamoto, H. Kohno and H. Fukuyama: $\JPSJ$ {\bf 63} (1994) 2741; 
	H. Fukuyama, H. Kohno, B. Normand and T. Tanamoto: {\rm J. Low
	Temp. Phys.} {\bf 99} (1995) 429.

\item{9)} B. Normand, H. Kohno and H. Fukuyama: $\PR$ {\bf B53} (1996) 856; 
$\JPSJ$ {\bf 64} (1995) 3903.

\item{10)} For a review see Proc. of the Yamada Conference XLI on Neutron
Scattering (Sendai, Japan; Oct., 1994), ed. by S. Funhashi, S. Katano
and R.A. Robinson (Elsevier Science Publishers B.V. and Yamada Science
Foundation, 1994).   

\item{11)} H. Yasuoka, T. Imai and T. Shimizu, in {\it Strong Correlations
and Superconductivity}, ed. by H. Fukuyama, S. Maekawa, and A.P. Malozemoff 
(Springer-Verlag, Berlin, 1989) 254.

\item{12)} N. Pyka {\it et al.},  $\PRL$ {\bf 70} (1993) 1457; M. K\"{a}ll
{\it et al.}, {\rm Physica} {\bf C 225} (1994) 317; H. Harashina  {\it et
al.}, preprint.

\item{13)} Y. Iye, in {\it Physical Properties of High Temperature
Superconductors III}, ed. by D.M. Ginsberg (World Scientific, 1991) 285.  

\item{14)} G. Baskaran and P.W. Anderson: $\PR$ {\bf B37} (1988) 580;
          G. Baskaran: Physica Scripta {\bf T27} (1988) 53.

\item{15)} N. Nagaosa and P.A. Lee: $\PRL$ {\bf 60} (1990) 2450;
          P.A. Lee and N. Nagaosa: $\PR$ {\bf B46} (1992) 5621.

\item{16)} R. Hlubina, W.O. Putikka, T.M. Rice and D.V. Khveshchenko: $\PR$
{\bf B46} (1992) 11224.

\item{17)} K. Kuboki: $\JPSJ$ {\bf 62} (1993) 420. 

\item{18)} M. Ubbens and P.A. Lee: $\PR$ {\bf B50} (1994) 438.

\item{19)} X.G. Wen and P.A. Lee: $\PRL$ {\bf 76} (1996) 503.

\item{20)} P.A. Lee, N. Nagaosa, T.K. Ng and X.G. Wen: unpublished.

\item{21)} Z.Y. Weng, D.N. Sheng and C.S. Ting: $\PR$ {\bf B49} (1994) 607; 
	$\PR$ {\bf B52} (1995) 637. 

\item{22)} V. Kalmeyer and R.B. Laughlin: $\PRL$ {\bf 59} (1987) 2095; 
       	R.B. Laughlin: $\PRL$ {\bf 60} (1988) 2677; Science {\bf 242}
	(1988) 525.

\item{23)} F. Basco, H. Kohno, H. Fukuyama  and G. Baskaran: $\JPSJ$ {\bf
	65} (1996) 687.

\item{24)} H. Mori: $\PR$ {\bf B46} (1992) 10952.

\item{25)} F.D.M. Haldane: $\PRL$ {\bf 67} (1991) 937.

\item{26)} C. Nayak and F. Wilczek: $\PRL$ {\bf 73} (1994) 2740.

\item{27)} Y. Hatsugai, M. Kohmoto, T. Koma and Y.-S. Wu: preprint. 

\item{28)} Y.-S. Wu: $\PRL$ {\bf 73} (1994) 922.

\item{29)} F.D.M. Haldane: $\PRL$ {\bf 66} (1991) 1529.

\item{30)} G. Baskaran: Int. J. Mod. Phys. Lett. {\bf B5} (1991) 643; See
also Hatsugai et. al. (Ref.27); M.D. Coutinho-Filho: private communication;
Y. Kato. and Y. Kuramoto:  $\JPSJ$ {\bf 65} (1996) 77; 1622. 

\item{31)} W. Chen and Y.J. Ng: $\PR$ {\bf B51} (1995) 14479.

\item{32)} M.V.N. Murthy and R. Shankar: $\PRL$ {\bf 72} (1994) 3629.

\item{33)} D. Yoshioka: $\JPSJ$ {\bf 58} (1989) 1516; D.P. Arovas and
A. Auerbach: $\PR$ {\bf B38} (1988) 316. 

\item{34)} T.M. Rice, in {\it The Physics and Chemistry of Oxide
Superconductiors}, ed. by Y. Iye, and H. Yasuoka (Springer-Verlag, Berlin,
1992), 313.

\item{35)} H. Fukuyama, Prog. Theor. Phys. Suppl. ${\bf 108}$ (1992) 287.

\item{36)} A.G. Loeser, Z.-X. Shen, D.S. Dessau, D.S. Marshall, C.H. Park,
P. Fournier, and A. Kapitulnik, Science {\bf 273} (1996) 325.

\item{37)} S. Deser, R. Jackiw, and S. Templeton: Annals of Physics {\bf
140} (1982) 372; $\PRL$ {\bf 48} (1982), 975.

\item{38)} W.H. Huang: $\PR$ {\bf E51} (1995) 3729.

\item{39)} J.B. Marston: $\PRL$ {\bf 61} (1988) 1914.




\vskip 5mm
\noindent
{\it Note added:} After submitting this paper, we found that
Huang$^{38)}$ gave an another description of $g$-on in terms of
ensemble of bosons whose fraction $g$ is randomly transmuted 
to fermions.  Based on his description an alternative formulation of our
theory is possible; gauge flux absorbed by spinons and holons is 
fluctuating discretely with values, 0 and $\p$,$^{39)}$ and thus their local
statistics is fluctuating between slave boson and slave fermion.
Note this picture applies to any dimensions.  This description leads to the
assignment of statistics defined as Case1 in the text.
\vskip 3mm

\vskip 5mm
\noindent
{\bf Figure Captions}
\vskip 3mm

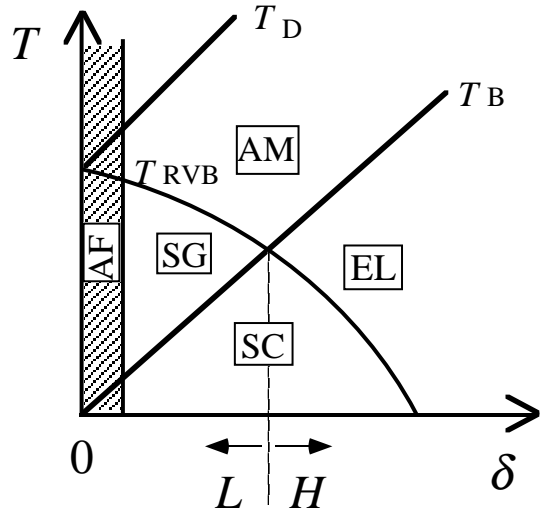
\item{{\rm Fig. 1.}} The slave boson mean-field phase diagram.  The phases
realized are antiferromagnetic (AF) state, superconducting (SC) state,
``spin gap'' (SG) phase, anomalous metallic (AM) state, and electron liquid
(EL) state.

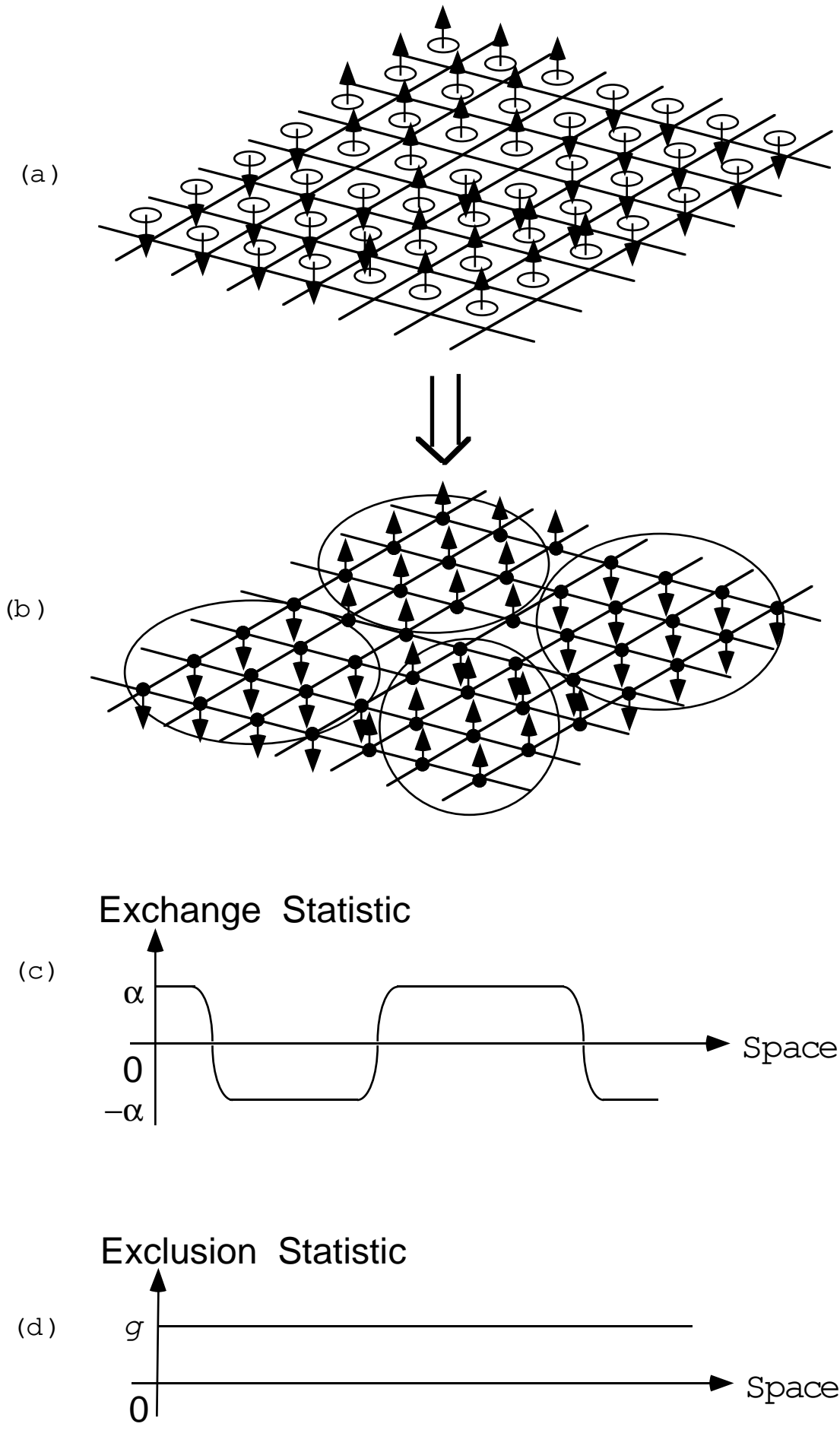
\item{{\rm Fig. 2.}} Temporal view of spatial variation of gauge flux (a),
its attachment to particles (b),  exchange statistics (c) and exclusion
statistics (d). 

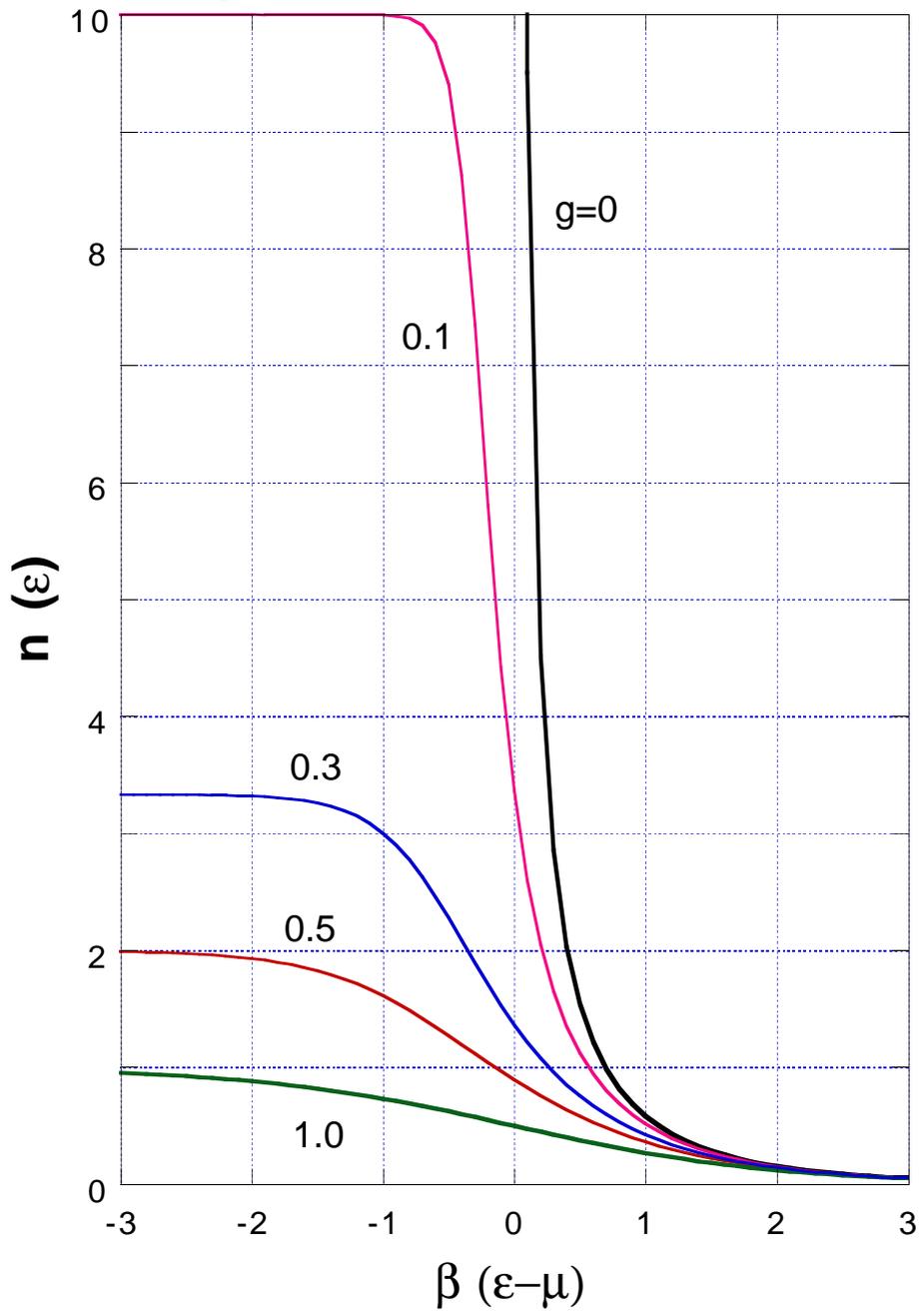
\item{{\rm Fig. 3.}} Distribution function of $g$-on, n($\e$), where  
 $\e$, $\mu$ and $\beta$ are energy, chemical potential and inverse
temperature, respectively.

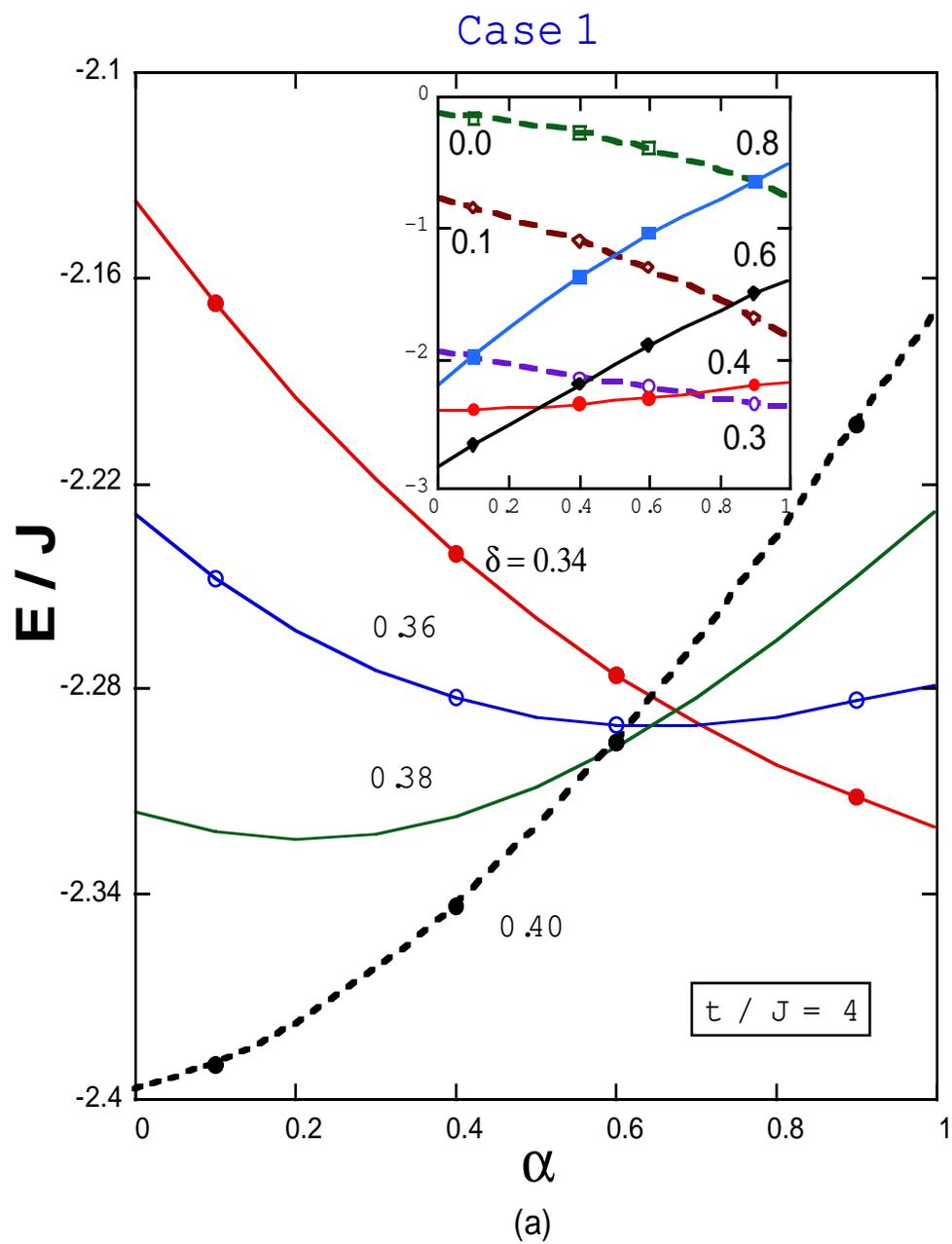
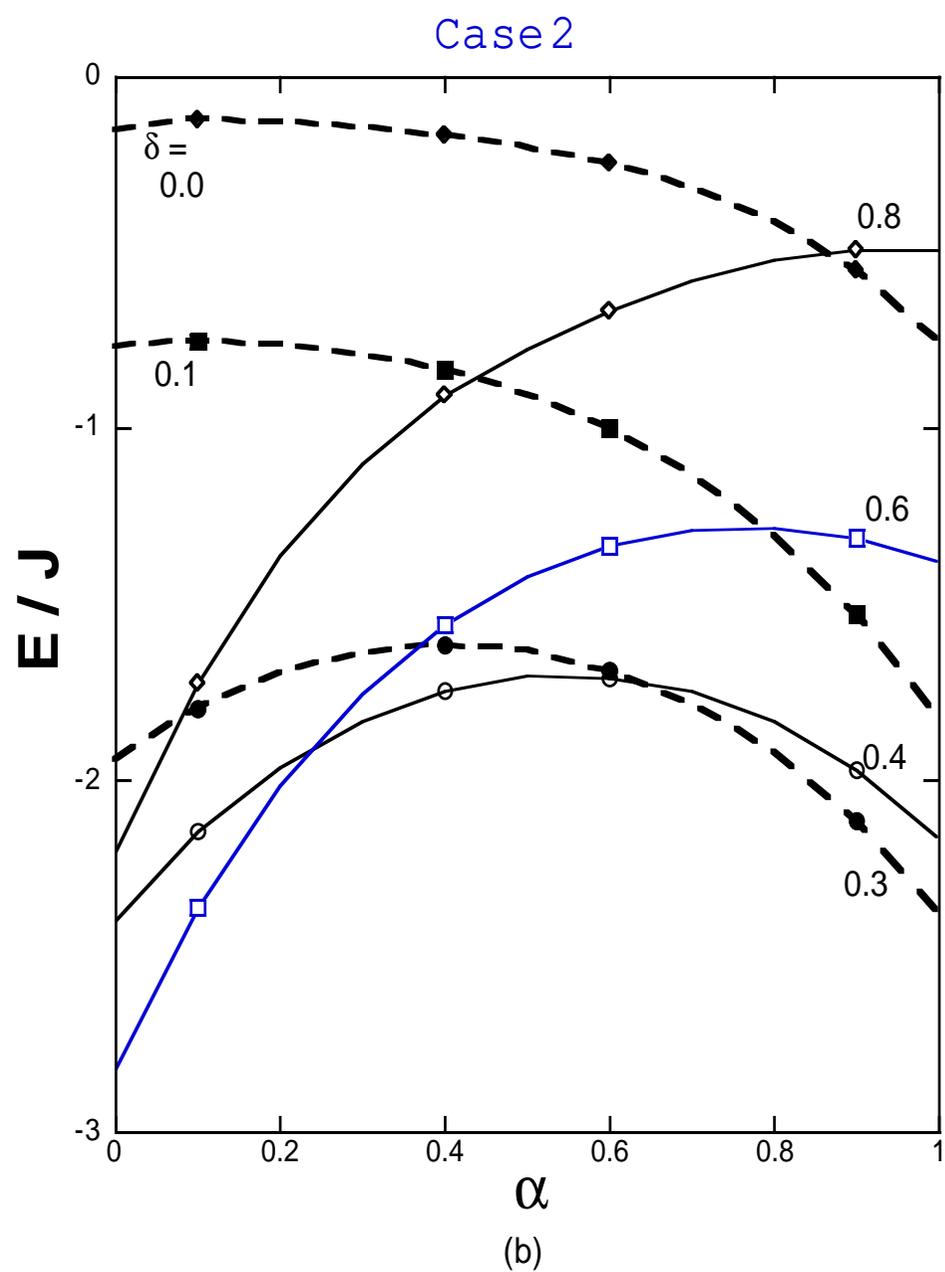
\item{{\rm Fig. 4.}}  The energy as a function of $\a$ for
several values of $\d$ for (a) Case1 and (b) Case2.  In Case1, the inset
shows those at the other values of $\d$. 

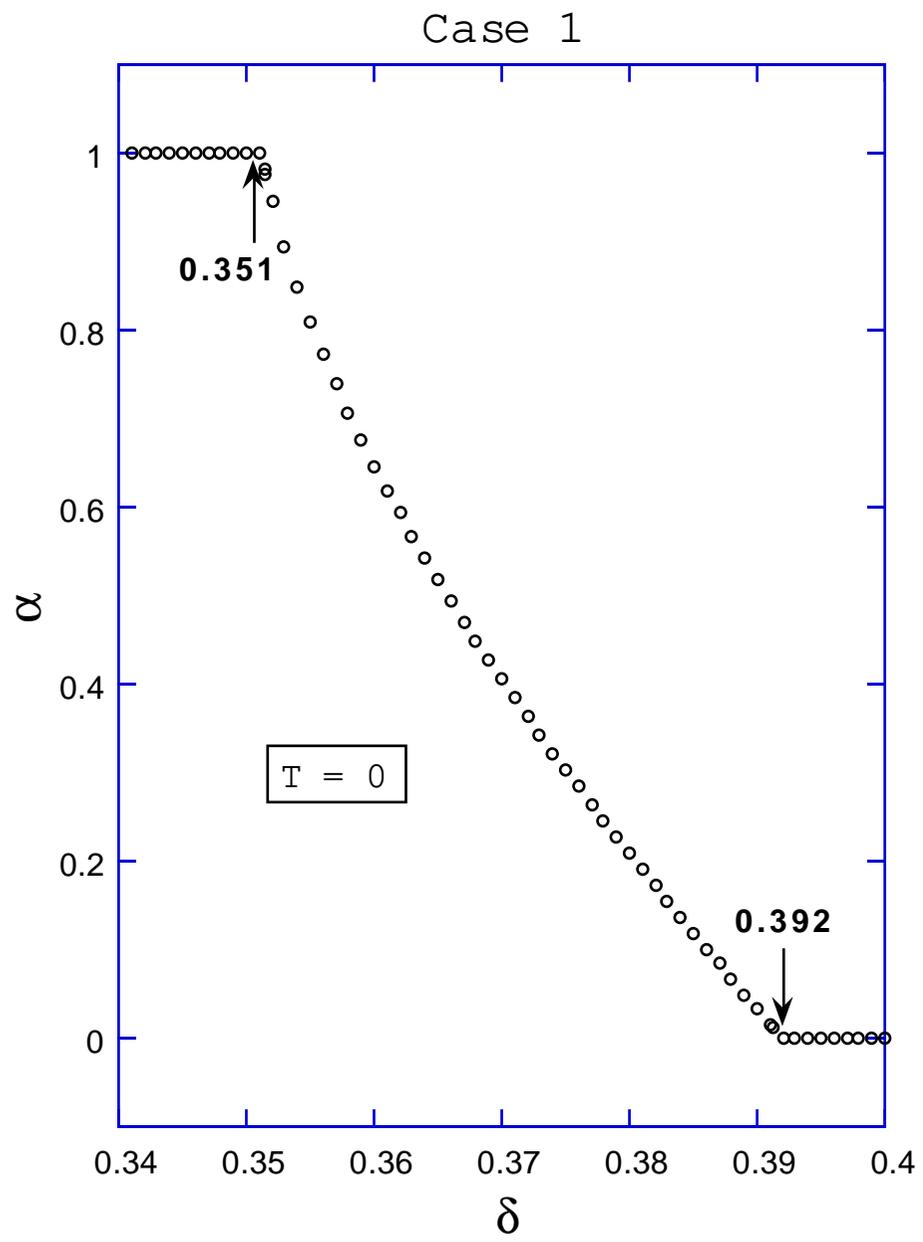
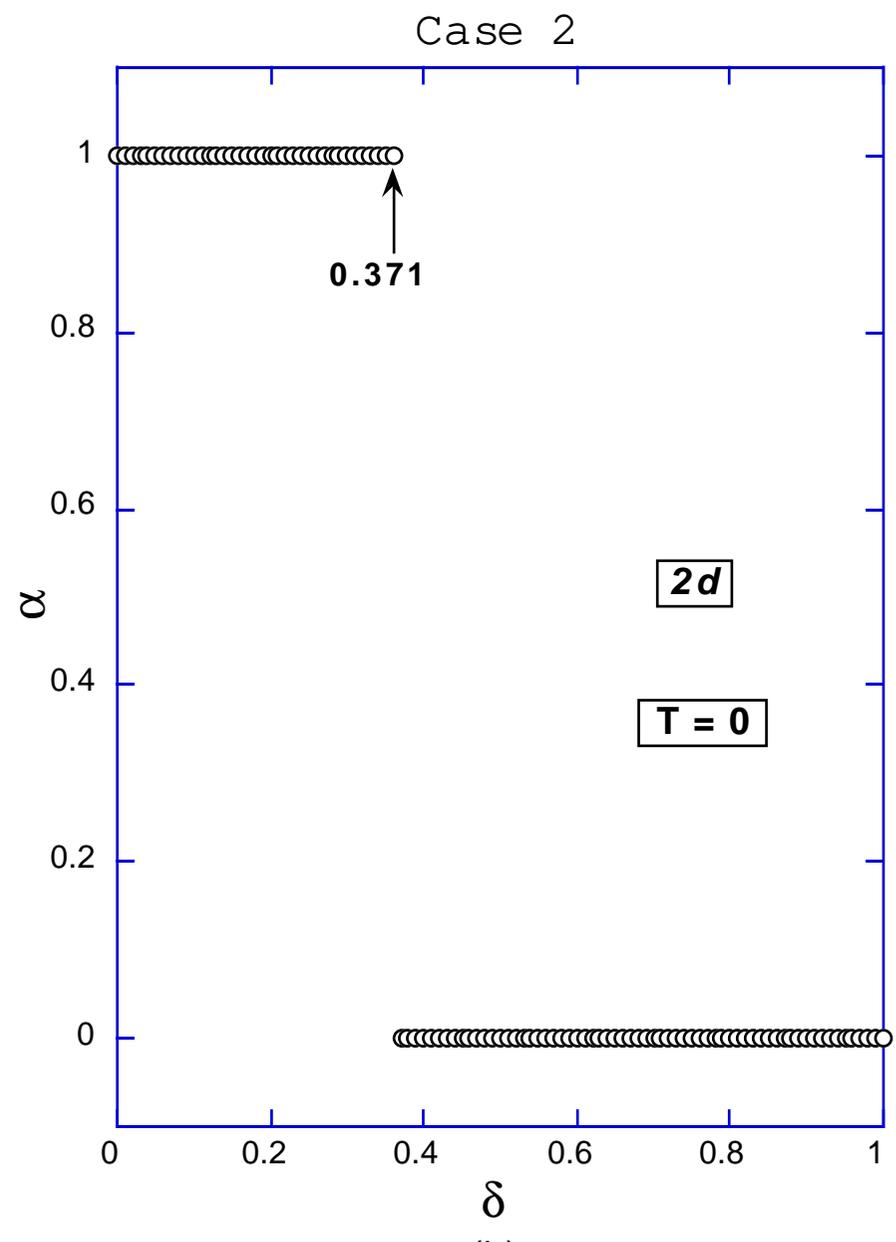
\item{{\rm Fig. 5.}}  The optimum values of $\a$ at $T=0$ as a function of $\d$
for (a) Case1 and (b) Case2.  

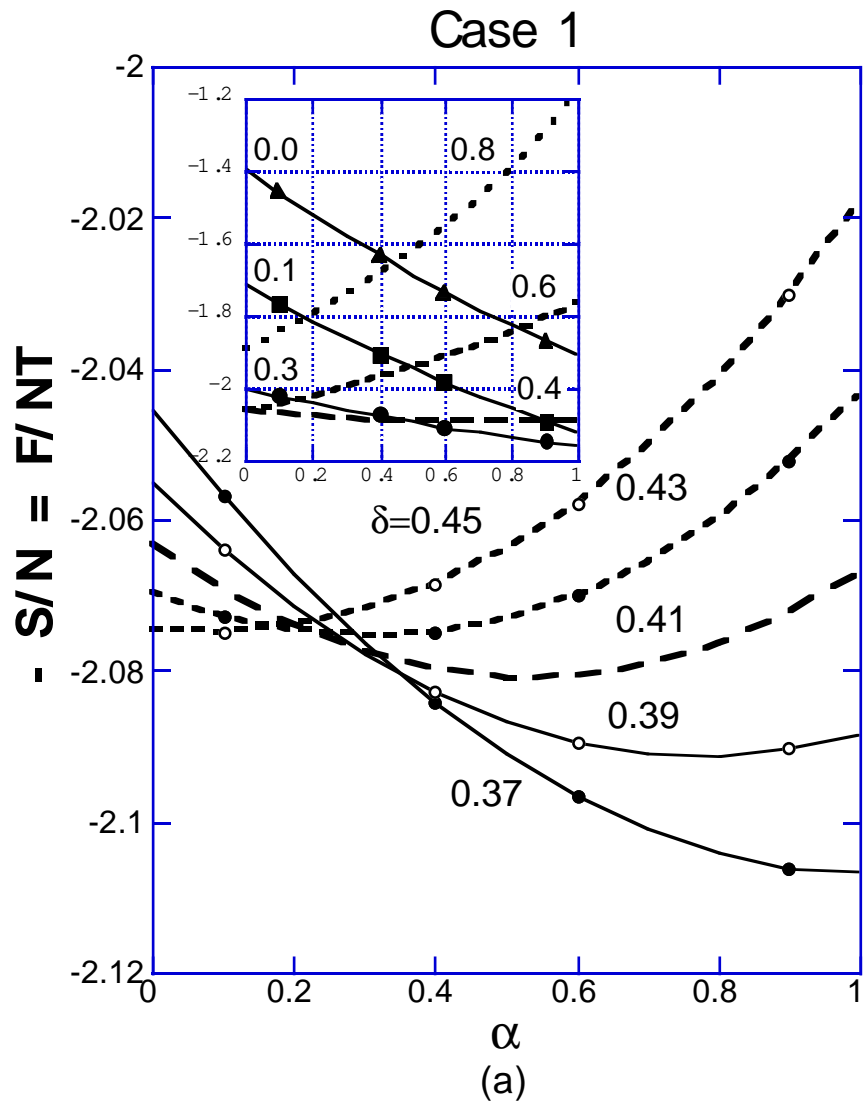
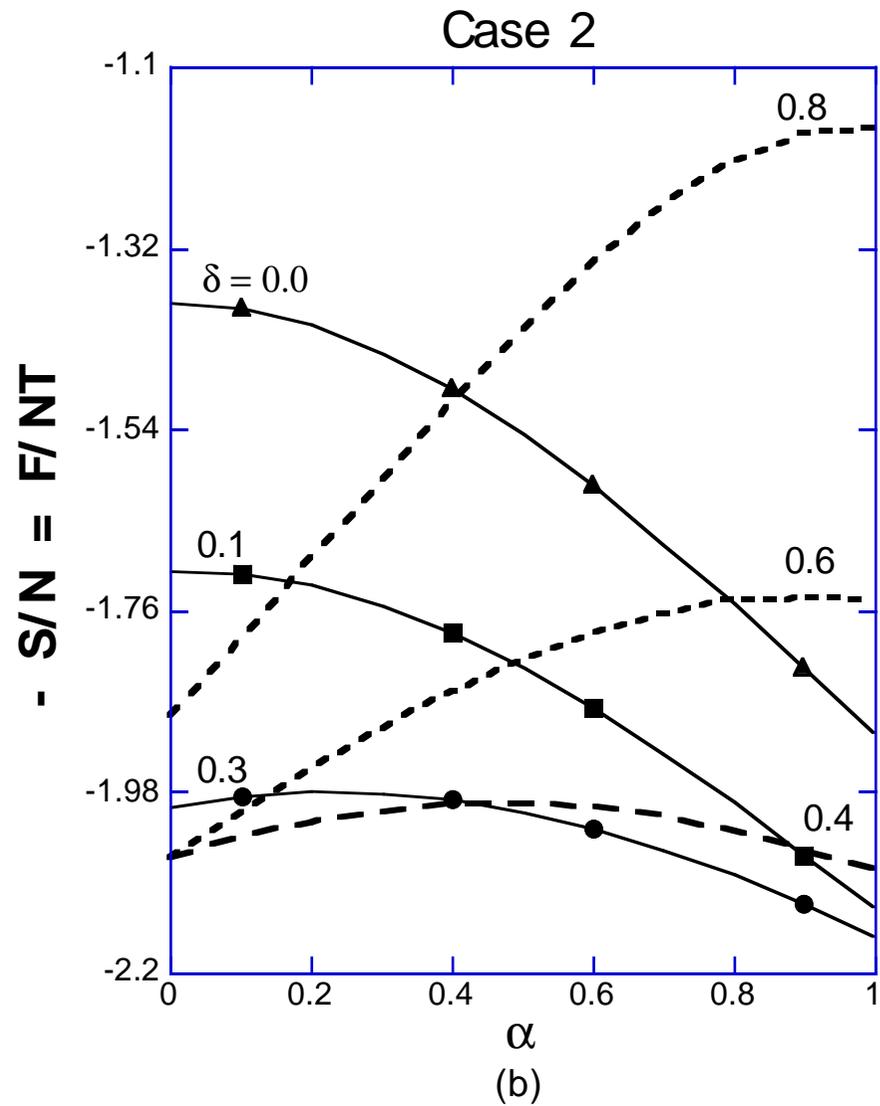
\item{{\rm Fig. 6.}}  The entropy for $T>T_D$ as a function of $\a$
for a choice of several values of $\d$ for (a) Case1 and (b) Case2.  These
are independent of temperature, spatial dimensionality and $J$.  

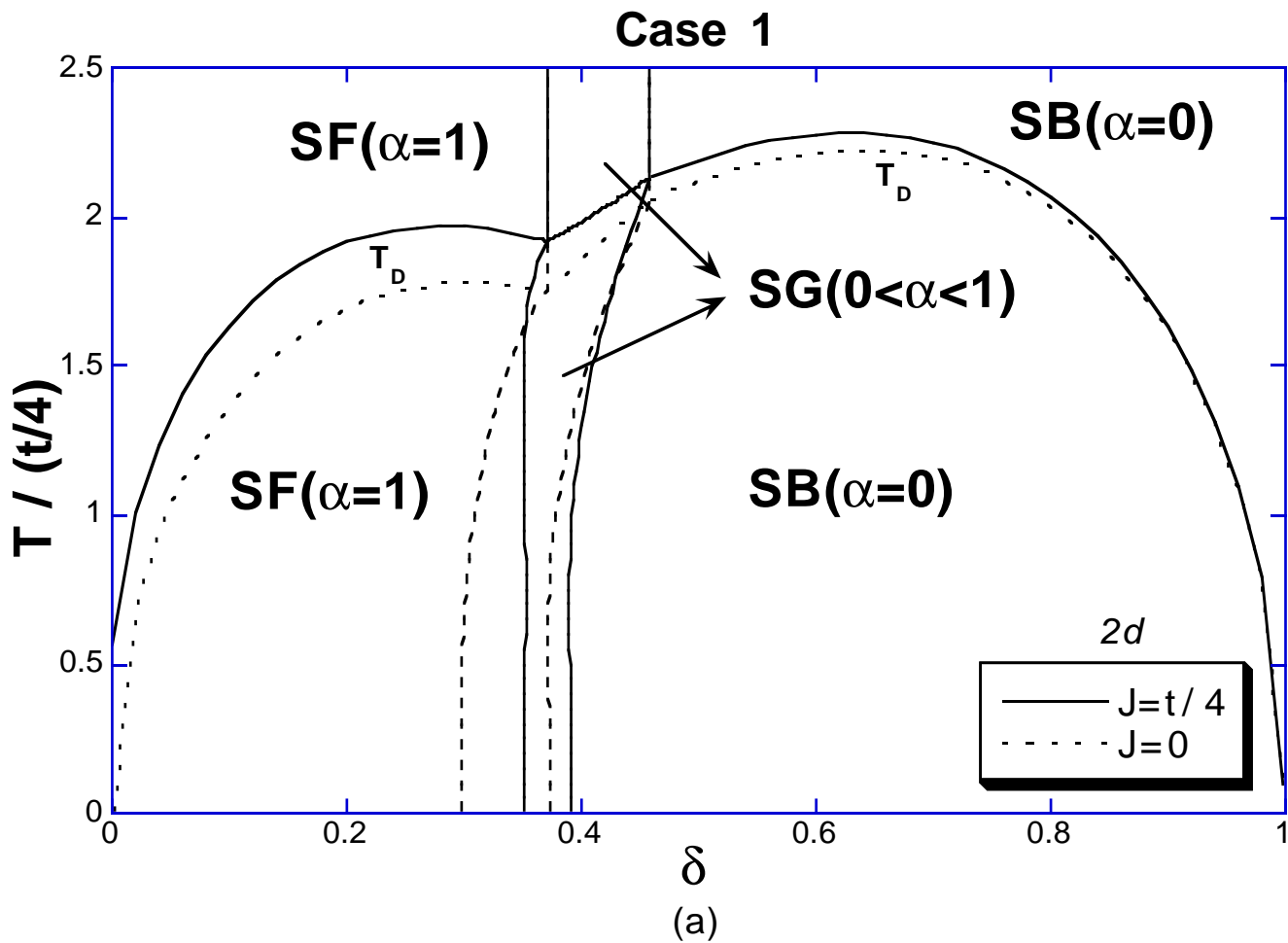
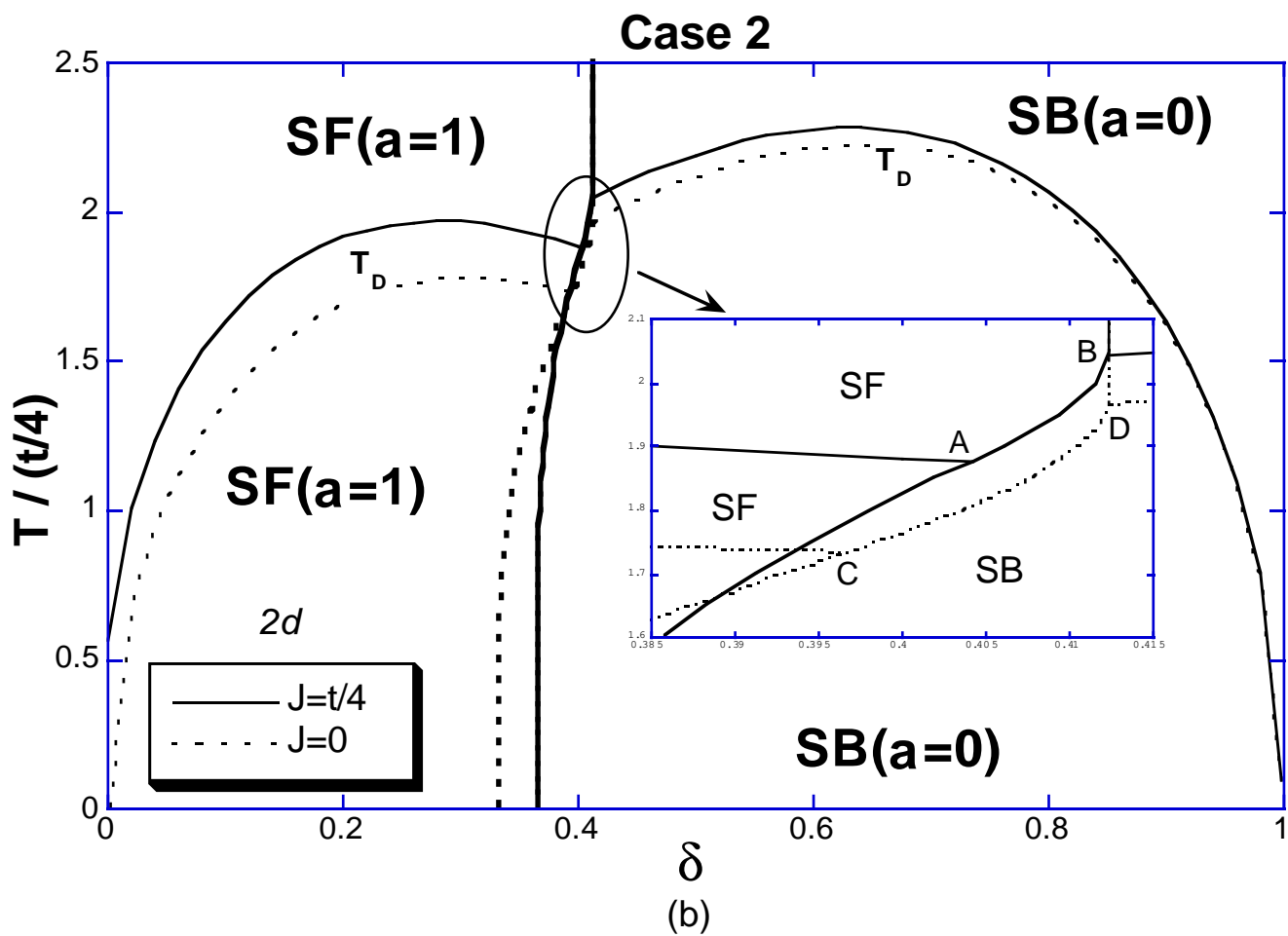
\item{{\rm Fig. 7.}}  The phase diagrams in $d=2$ for (a) Case1 and (b)
Case2.  The inset in (b) shows the region where the transition to $u$-RVB
state is of first order (from $A$($C$) to $B$($D$) for $J=t/4$ ($J=0$)).  The
solid and dashed lines are for $J=t/4$ and $J=0$, respectively. 

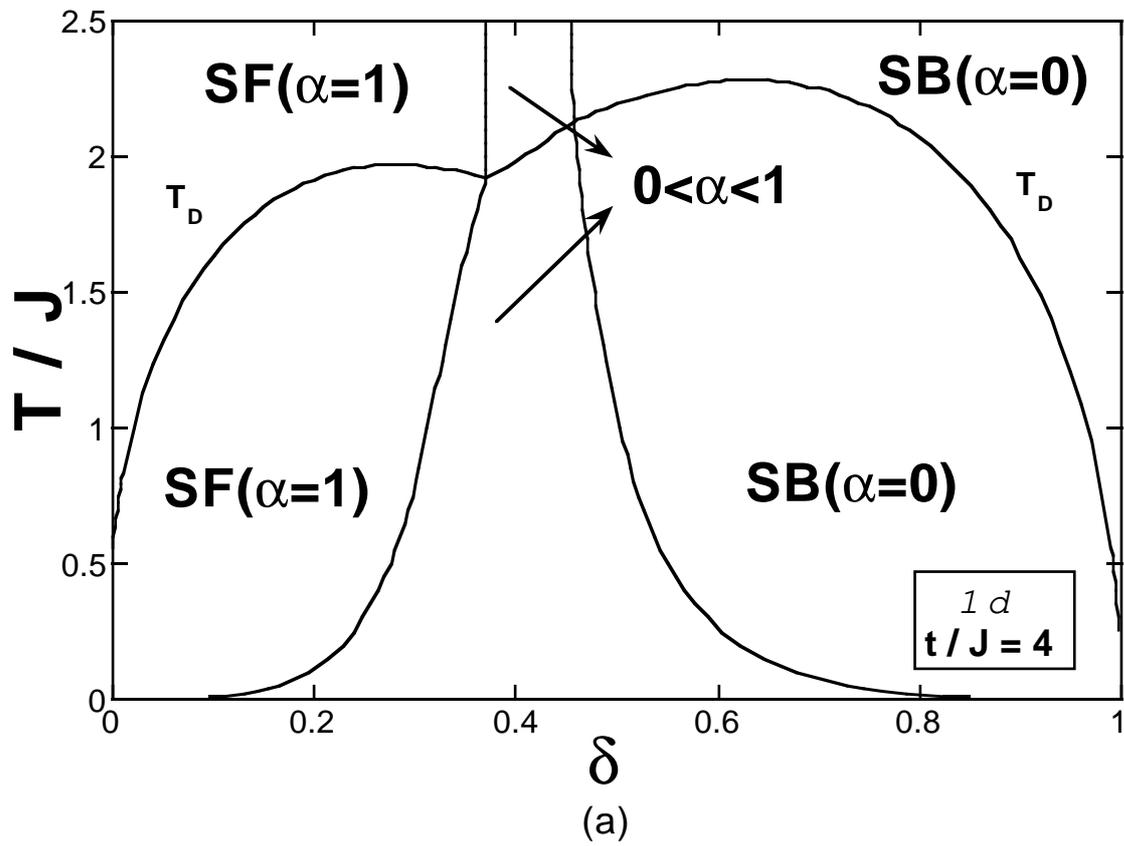
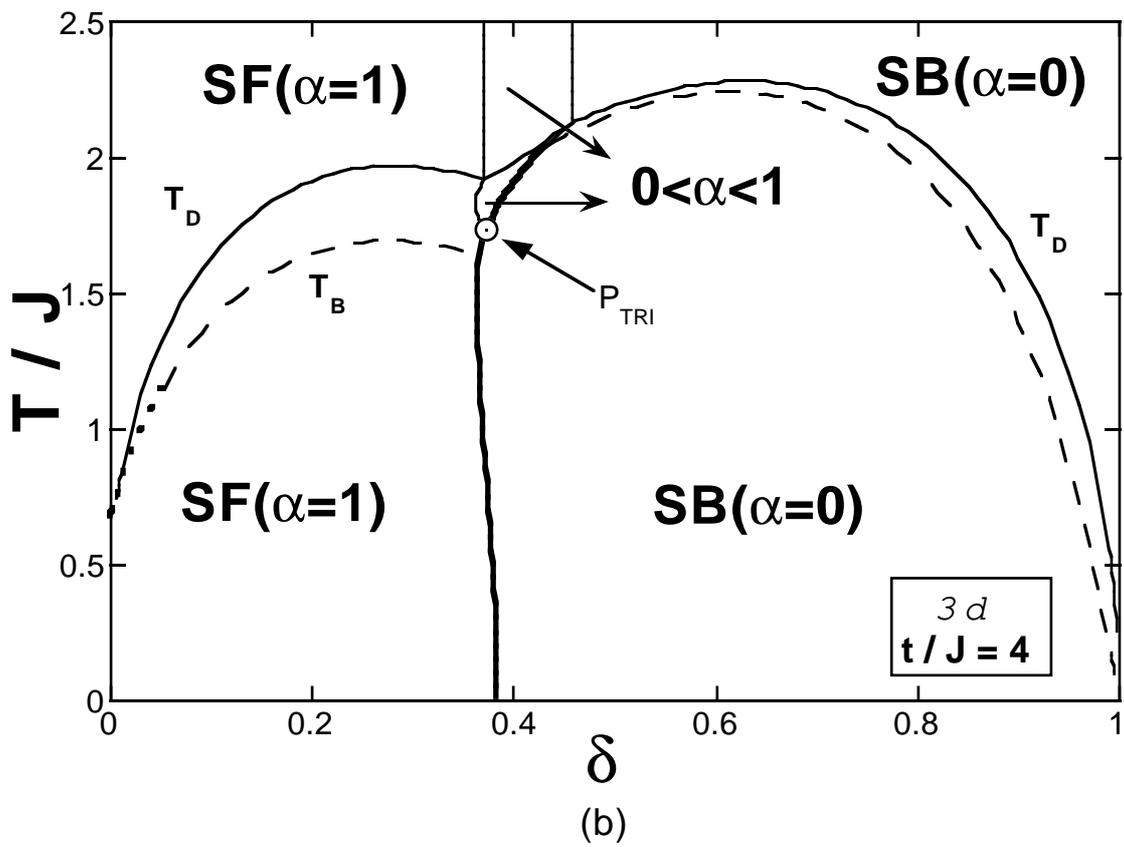
\item{{\rm Fig. 8.}}  The phase diagrams for (a) $1d$ and (b) $3d$.  In
(b), the dotted line in $T_B$ and in $T_D$ represents first order
transitions.   (c) The enlarged version of (b)
where slave $g$-on phase is realized.  The inset shows the $\a$-dependence  
of the free energy measured relative to slave fermion or slave boson on points
X, Y, Z.  (d) The doping dependences of the optimum statistics are shown for
$d=1, 2, 3$ at $T=0$.    

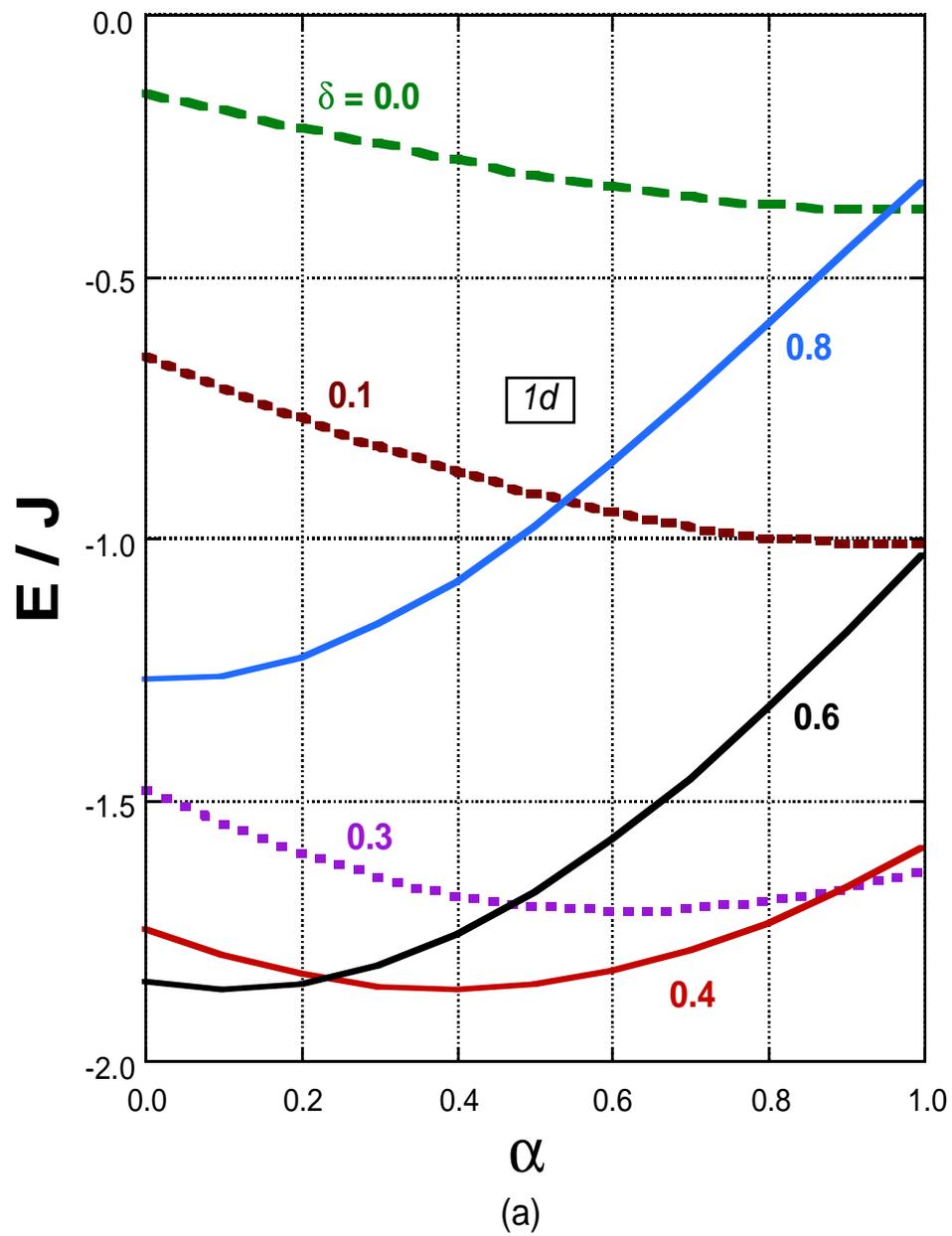
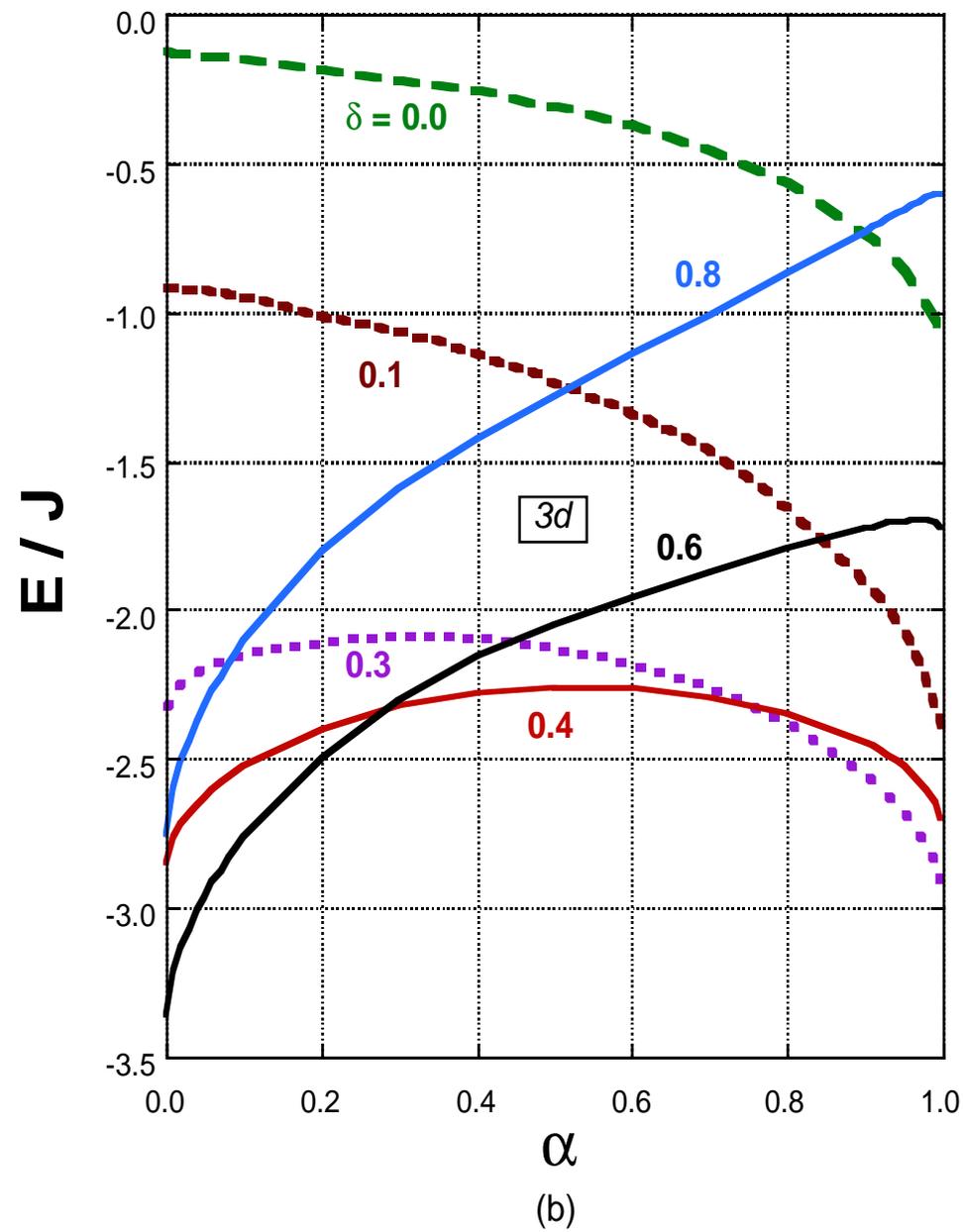
\item{{\rm Fig. 9.}}  The $\a$-dependence of energy for several values of $\d$ 
in (a) $1d$ and (b) $3d$ at $T=0$.

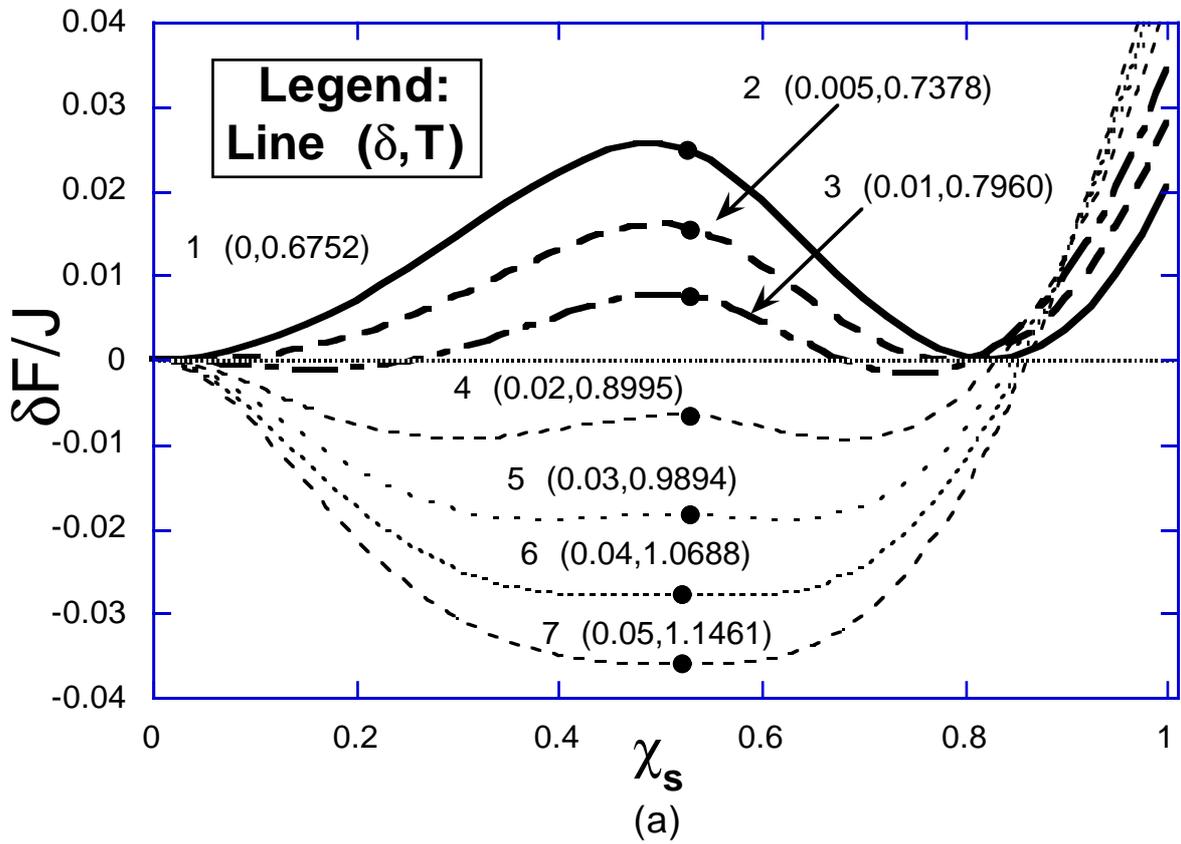
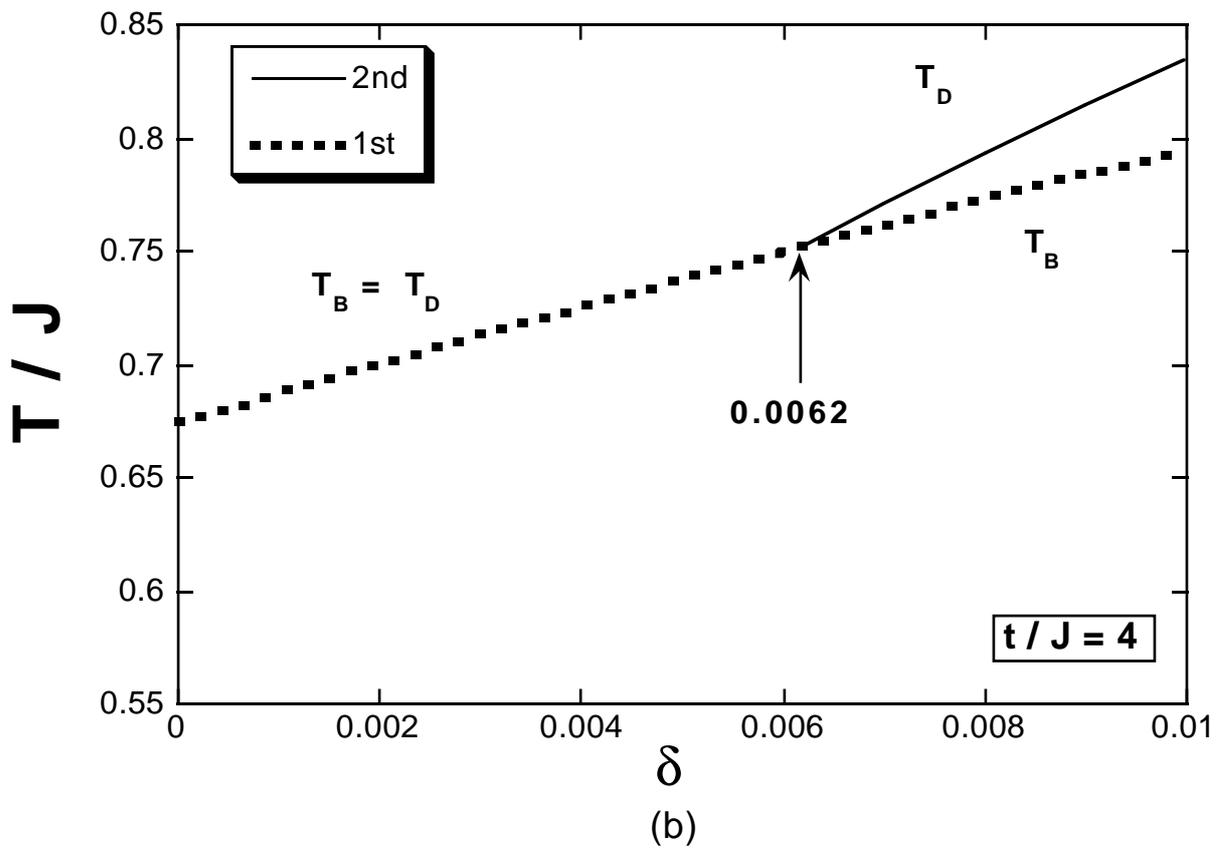
\item{{\rm Fig.10.}}  (a)  The Ginzburg-Landau free energy as a
function of $u$-RVB order parameter, $\chi_s$, close to half-filling at
temperature chosen such that the two minima have the same value.  At the
dot mark on each line bose condensation sets in.  Lines 1-2 show why $T_D$
coincides with $T_B$ and becomes of first order transition.  The series of
lines 1-7 shows how the two minima of $\chi_s$ smoothly merges into single
point, i.e. $T_B$ reduces of second order as $\d$ increases.  (b)  The
$\d$-$T$ phase diagram or small values of $\d$ where the $T_D$ coincides
with $T_B$.  The transition  in $T_D$ changes from first order to second
order t $\d=0.0062$.

\vskip 5mm
\noindent

\bye